\documentclass[manuscript]{aastex}

\shorttitle{Flares at Solar Hale Boundaries}
\shortauthors{Svalgaard et al.}

\begin{document}
\title{Flaring Solar Hale Sector Boundaries
}

\author{L. Svalgaard}
\affil{HEPL, Stanford University, Stanford, CA 94304}
\email{leif@leif.org}
\and
\author{I. G. Hannah}
\affil{School of Physics \& Astronomy, University of Glasgow, Glasgow, Scotland, G12 8QQ}

\begin{abstract}
Magnetic fields and the occurrence of flares and microflares are strongly concentrated 
near that portion (the Hale boundary) in each solar hemisphere where the change in 
magnetic sector polarity is the same as that between leading and following sunspot 
polarities in active regions.
\end{abstract}

\keywords{Sun: Hale sector polarity, Flare distribution}

\section{Introduction}
When the Heliospheric Current Sheet overtakes the Earth (or a spacecraft) 
an abrupt change of magnetic polarity (away from the Sun or towards the Sun) is 
observed: a Sector Boundary (SB). 
 
A SB (observed at Earth) maps back to a magnetic neutral line at central meridian in the 
solar corona and also in the originating photospheric magnetic field about 5 days earlier, 
being the transit time of the solar wind. The latter neutral line is on average meridional,
 i.e. along longitudes \citep{SvaXxx75}. 

\begin{figure}
\centerline{\plotone{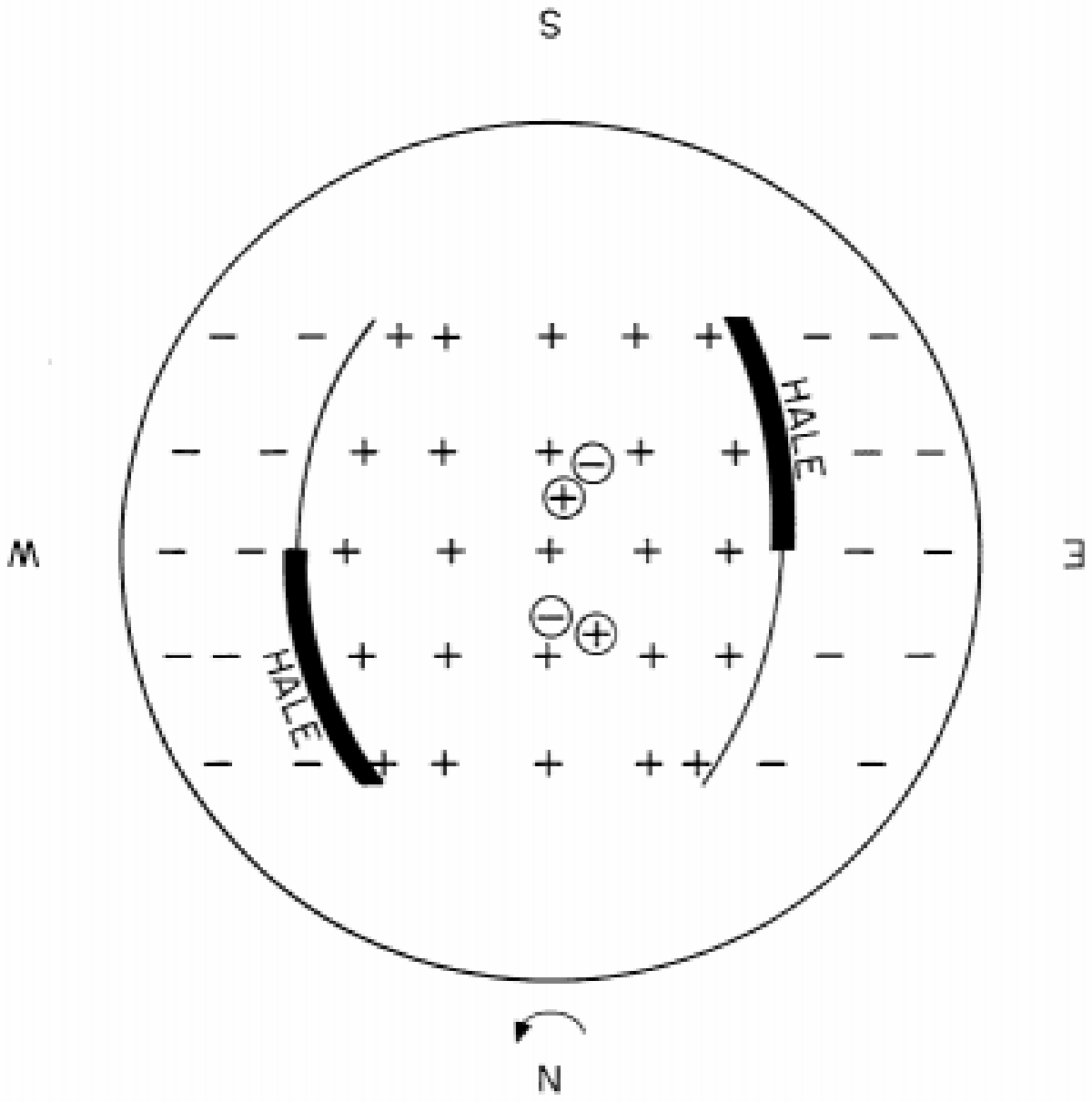}}
\caption{Schematic of the solar disk showing the portion of a sector boundary that is 
designated a Hale boundary, {\it i.e.} that portion of a sector boundary that is located 
in the solar hemisphere in which the change of magnetic polarity across the sector 
boundary is the same as the change of magnetic polarity from a preceding spot to a 
following spot. The spot polarities shown in the small circles correspond to 
even-numbered cycles, {\it e.g.} cycle 24. \protect\citep{SvaWil76}. 
} \label{F-HaleB}
\end{figure}

Active regions have a photospheric neutral line too. The neutral line divides opposite 
polarities oppositely in opposite hemispheres (Hale's law). The portion (in a hemisphere, 
see Figure~\ref{F-HaleB}) of a SB neutral line where the polarity change is the same as 
that for an active region was called a Hale Boundary by \cite{SvaWil76}.

They showed that above a Hale portion of a SB, the green corona has a maximum in 
brightness, while above a non-Hale boundary, the green corona has minimum brightness. 
Using synoptic maps of the magnitude of the photospheric field strength observed at 
Mt. Wilson Observatory during 1967 to 1973 it was also found that the unsigned 
magnetic field flux is at a maximum at the Hale boundary, in concert with the green 
corona brightness.

\section{Observations}
\subsection {Magnetic Field}
We have recently repeated the analysis using all available data from the Wilcox Solar 
Observatory (WSO). At WSO (\url{http://wso.stanford.edu/}) magnetograms using the 
525 nm Fe \textsc{i} line are obtained every day with a sufficiently clear sky. Conditions 
permitting, several magnetograms may be secured on a given day. Observational details 
can be found elsewhere: \cite{SchXxx77,SvaXxx78,DuvXxx78}. The resulting magnetogram 
is a $21\times21$ array oriented north-south on the Sun and has not been remapped to any 
other coordinate system. In the analysis we ignore the annual variation of latitude of disk 
center, giving rise to less than 1\% effect on the measured field \citep{DuvXxx78}. 
The magnetograms show the line-of-sight magnetic flux density over the 3' aperture, 
not corrected for magnetograph saturation.

\begin{figure}
\centerline{\plotone{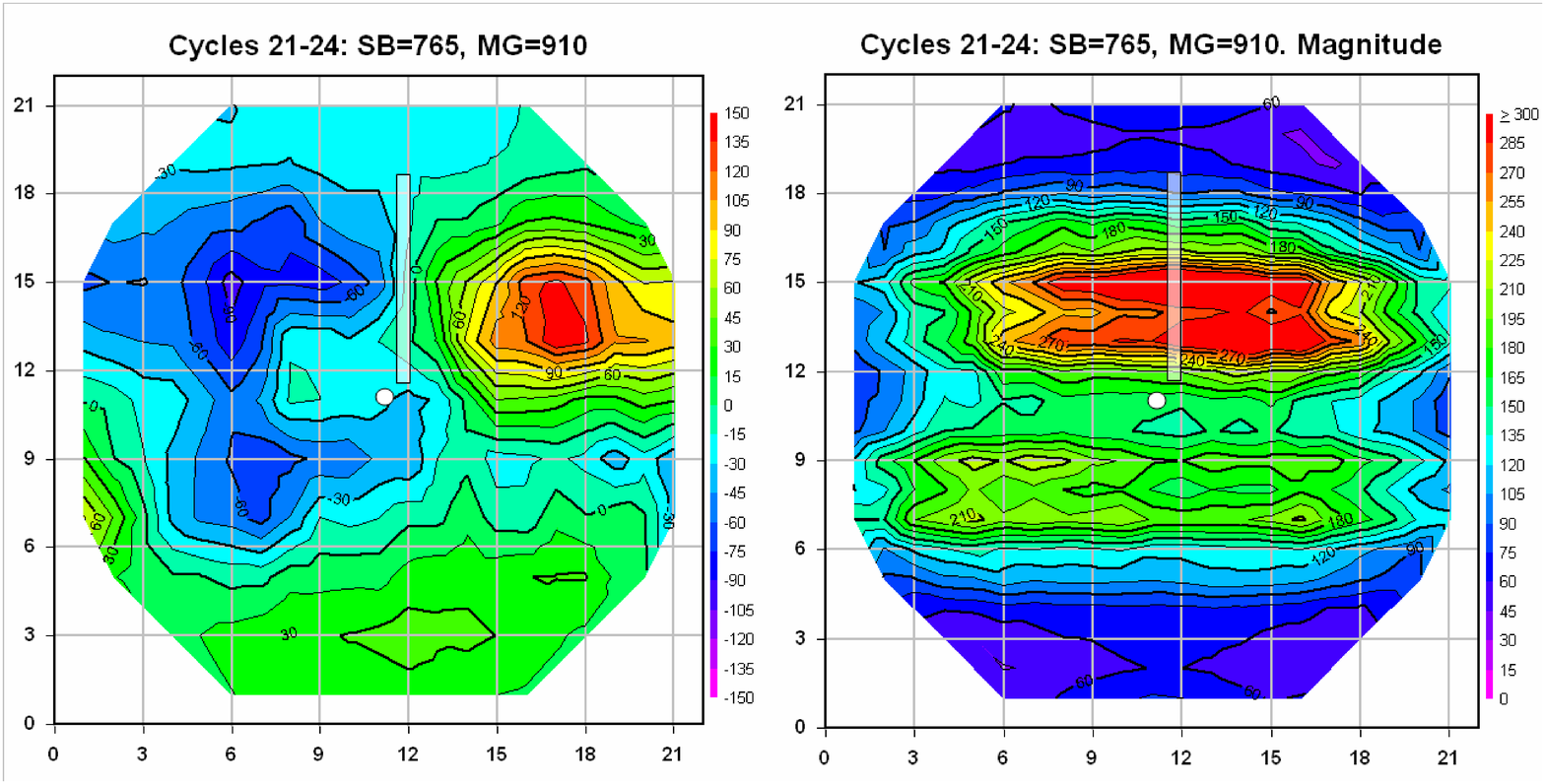}}
\caption{The average magnetograms for a nominal (+,-) Hale boundary during solar cycle 23. 
A total of 910 magnetograms were superposed on 765 SBs (mapped back on the Sun) using 
the WSO observations 1976-2010. Data has appropriately been mirrored and sign-reversed 
as described in the text to reduce all data to the situation for cycle 23. The Hale portion of 
the sector boundary is marked by the semi-transparent bar. Left Figure shows the average 
signed field, while the right-hand Figure shows the average unsigned (total) field.
} \label{F-HaleM}
\end{figure}

We superpose full-disk magnetograms  from the times where a SB was at [i.e. within a day] 
central meridian. We can take advantage of the polarity changes and treat (-,+) boundaries 
as (+,-) boundaries by reversing the sign of the magnetic field, and of the Hale magnetic 
cycle by reversing latitudes between cycles to construct a ‘nominal’ cycle 23 magnetogram 
for a (+,-) Hale boundary from the average over solar cycles 21-24 shown in Figure 
\ref{F-HaleM}.

It is now evident, that on average, what causes the ‘warps’ in the current sheet (and
hence the SB at Earth) originates strongly from one hemisphere, namely that which 
has the Hale boundary (the Northern in Figure \ref{F-HaleM}). Coronal holes are often 
found co-located with the interiors of the sectors.

One important (the only?) source of the open flux is dispersing flux from active regions 
with strong magnetic fields so we expect the concentration of total flux at the Hale 
boundary as shown in the right-hand side of Figure \ref{F-HaleM}. The closed field lines 
associated with the active regions would trap coronal material, explaining the enhanced 
brightness of the green corona at Hale boundaries.

\subsection {Flares}
We would also expect flares and microflares to preferentially occur near the Hale boundary. 
Using the RHESSI list of hard X-ray flares covering the interval March 2002 to March 2008 
(wholly within cycle 23) we find, indeed, that to be the case, Figure \ref{F-HaleF}. The green 
boxes show where flares are expected, based on association with strong magnetic fields: 
i.e. at the Hale boundary. The red circles show that hardly any flares occur near a non-Hale 
boundary. 

\begin{figure}
\centerline{\plotone{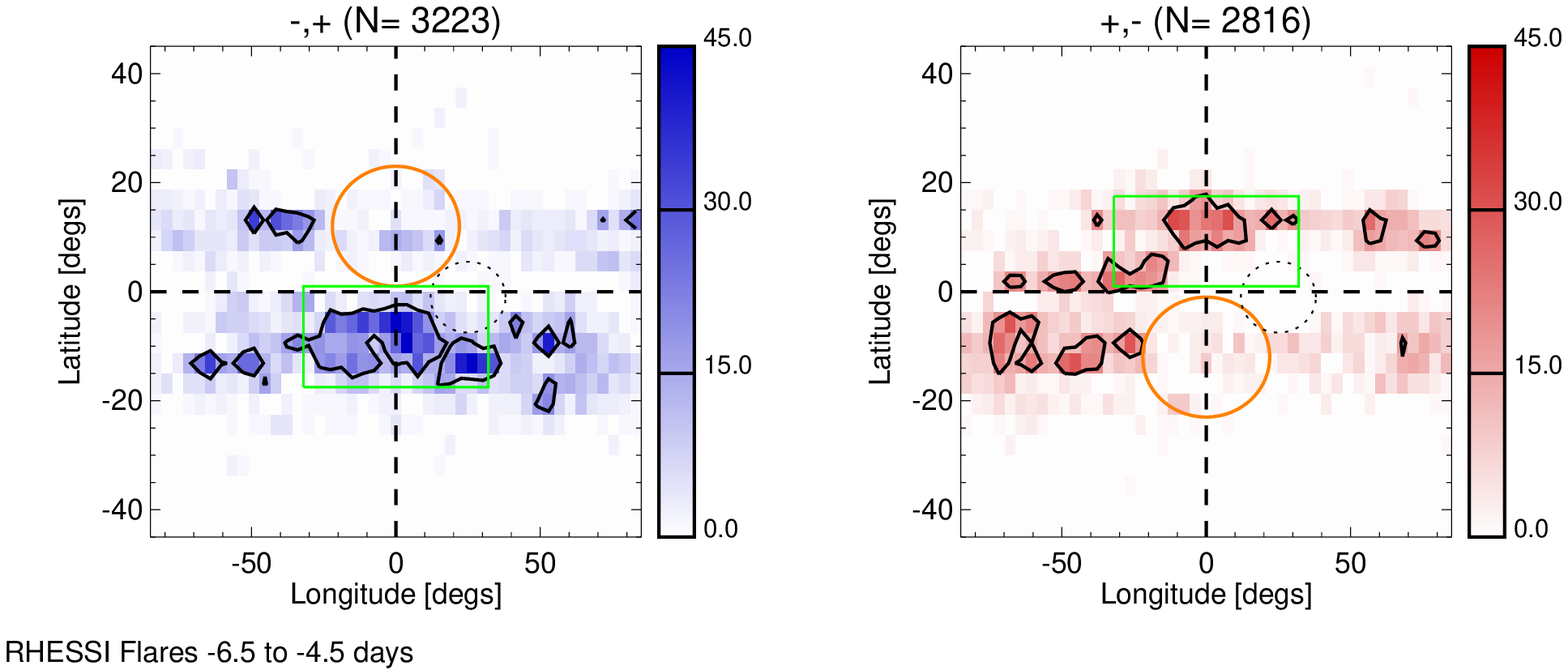}}
\caption{Distribution of RHESSI flares within $\pm24$ hours of 223 sector boundaries mapped 
back to central meridian (dashed vertical line) for part of solar cycle 23, March 2002 to March 2008. 
The green boxes show where flares are expected, based on association with strong magnetic 
fields: i.e. at the Hale boundary. The red circles show that hardly any flares occur near a non-Hale 
boundary. The number of flares in each distribution is shown above each plot. Only flares within 
$\pm85^\circ$ of CM are counted. The small dashed line circles show the imaging axis bias area.
} \label{F-HaleR}
\end{figure}

The RHESSI flare list catalogues times of sharp rises in X-ray flux detected between 3 to 
50 keV. Using the imaging capabilities of RHESSI the position of each of these events is 
found to arc second accuracy. The RHESSI flare list covers the largest GOES $>$ X10-class 
flares down to A-Class microflares ($> 10^{-3}$ to $10^{-8}$ Wm$^{-2}$ in terms of GOES 
1-8 \AA \hspace{1pt} flux). An instrumental bias in the RHESSI event list is a deficiency of 
events about the location of spacecraft pointing, predominantly slightly West of disk center 
(dashed circles on Figure \ref{F-HaleR}; see also Figure 4 of \cite{ChrXxx08}. RHESSI's 
imaging method uses rotation modulation collimators, so flares close to where this imaging 
axis is pointing will have their location poorly determined and are not included in the flare 
list. For the RHESSI microflare study it was estimated that less than 2\% of events were 
affected by the imaging axis problem \citep{ChrXxx08}.

\begin{figure}
\centerline{\plotone{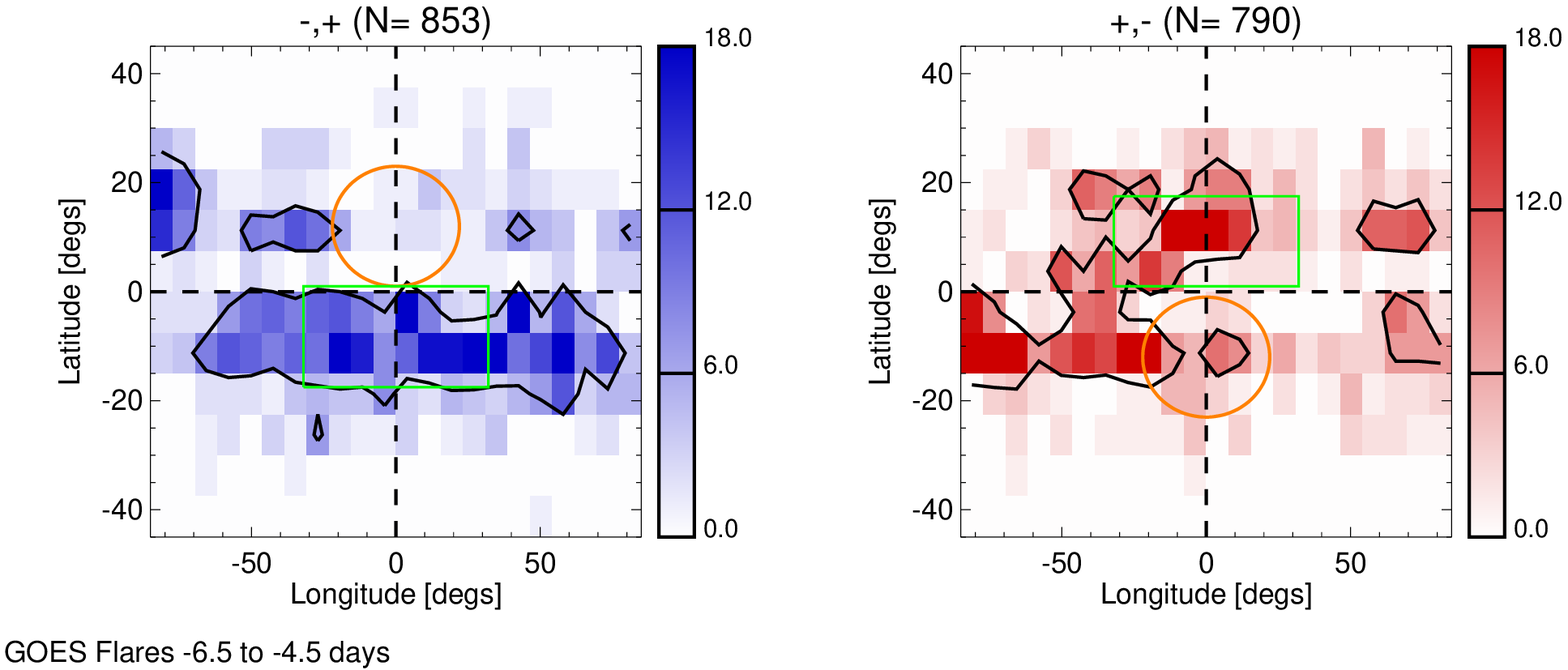}}
\caption{Distribution of GOES flares within $\pm24$ hours of 392 sector boundaries mapped 
back to central meridian (dashed vertical line) for solar cycle 23, May 1996 through 2008. 
The green boxes show where flares are expected, based on association with strong magnetic 
fields: i.e. at the Hale boundary. The red circles show that hardly any flares occur near a non-Hale 
boundary. The number of flares in each distribution is shown above each plot. Only flares within 
$\pm85^\circ$ of CM are counted. GOES data does not have the same imaging bias as RHESSI.
} \label{F-HaleG}
\end{figure}

GOES flares for May 1996 through 2008 (cycle 23) show the same distribution, Figure 
\ref{F-HaleG}. The GOES flare list contains events down to the B1-class, $10^{-7}$ Wm$^{-2}$, 
about an order of magnitude larger than the smallest events in the RHESSI flare list. 

An early analysis by \cite{Ditt75} also concluded that flares preferentially occur near sector 
boundaries whose polarity agrees with that of bipolar active regions a given by the Hale 
polarity laws. Already \cite{BumObr69} found that flares and especially `proton'-flares tend 
to occur near SBs, and \cite{GriXxx86} found that toroidal magnetic flux emerges preferably 
at Hale boundaries.

\section {Conclusion}
In a sense this is just what we would expect. On the other hand, the patterns we report 
emphasize the high degree of coherence in the organization of solar magnetic activity on 
large scales, something that may not be well-understood theoretically, but which 
presumably links the sector structure to the deep interior of the Sun. The solar sector 
structure is very organized and long-lived and it seems that solar activity enjoys some 
of the same degree of spatial and temporal structure. Such structure might be useful for 
prediction of flares.

\acknowledgments
Acknowledgments:
We are grateful to the staff of the Wilcox Solar Observatory and to the funding
agencies that have made this exceptional dataset possible. Hugh Hudson was instrumental
in getting the research for the present paper started. WSO magnetograms are available 
on request from \url{http://wso.stanford.edu/}. The sector boundary list can be downloaded 
from the WSO site at \url{http://wso.stanford.edu/SB/SB.Svalgaard.html}. The RHESSI flare 
list can be found at \url{http://hesperia.gsfc.nasa.gov/hessidata/dbase/hessi\_flare\_list.txt}.

{\it Facilities:} \facility{WSO}, \facility{RHESSI}.

\clearpage

\begin{thebibliography}{}

\bibitem[Bumba \& Obridko(1969)]{BumObr69} Bumba, V., \& Obridko, V. N. 1969, 
Solar Phys., 6, 104

\bibitem[Christie, Hannah, Krucker, McTiernan, \& Lin(2008)]{ChrXxx08} Christie, S., Hannah, 
I. G., Krucker, S., McTiernan, J., \& Lin, R. P. 2008, Ap. J., 677, 1385

\bibitem[Dittmer(1975)]{Ditt75} Dittmer, P. H. 1975, Solar Phys., 41, 227

\bibitem[Duvall, Scherrer, Svalgaard, \& Wilcox(1978)]{DuvXxx78} Duvall, T. L. Jr., Scherrer, P. H., 
Svalgaard, L., \& Wilcox, J. M. 1978, Solar Phys., 61, 233

\bibitem[Grigoryev, Latushko, \& Peshcherov(1986)]{GriXxx86} Grigoryev, V. M., 
Latushko, S. M., \& Peshcherov, V. S. 1986, Contrib. Astron. Obs. Skalnate Pleso, 15, 481

\bibitem[Scherrer, Wilcox, Svalgaard, Duvall, Dittmer, \& Gustavson(1977)]{SchXxx77} Scherrer, 
P. H., Wilcox, J. M., Svalgaard, L., Dittmer, P. H., \& Gustavson, E. K. 1977, Solar Phys., 54, 353

\bibitem[Svalgaard, Duvall, Scherrer(1978)]{SvaXxx78} Svalgaard, L., Duvall, T.L.Jr., \& Scherrer, 
P.H. 1978, Solar Phys., 58, 225

\bibitem[Svalgaard, Wilcox, Scherrer, \& Howard(1975)]{SvaXxx75} Svalgaard, L., Wilcox, J. M., 
Scherrer, P. H., \& Howard, R. 1975, Solar Phys., 45, 83

\bibitem[Svalgaard \& Wilcox(1976)]{SvaWil76} Svalgaard, L., \& Wilcox, J. M. 1976, 
Solar Phys., 49, 177

\end{thebibliography}
\end{document}